\addtocontents{toc}{\protect} 
\documentclass[preprint]{elsarticle}
\usepackage[left=3cm,right=3cm,top=3cm,bottom=3.5cm]{geometry}
\usepackage{amsmath}
\usepackage[utf8]{inputenc}
\usepackage[english]{babel}

\usepackage{lineno,hyperref}
\modulolinenumbers[5]

\usepackage{graphicx}
\usepackage{subcaption}
\usepackage{float}
\usepackage{natbib}
\usepackage{booktabs}
\usepackage{cleveref} 
\usepackage{amsfonts}

\newcommand{\sref}[1]{Sec.~\ref{#1}}
\newcommand{\fref}[1]{Fig.~\ref{#1}}

\newcommand{\tref}[1]{Tab.~\ref{#1}}
\newcommand{\eref}[1]{Eq.~\ref{#1}}

\newcommand{\doi}[1]{\href{http://dx.doi.org/#1}{\texttt{doi:#1}}}

\makeatletter
\def\ps@pprintTitle{%
  \let\@oddhead\@empty
  \let\@evenhead\@empty
  \def\@oddfoot{\reset@font\hfil\thepage\hfil}
  \let\@evenfoot\@oddfoot
}
\makeatother

\newpage
\title{Propagation of controlled frontward impulses through standing crowds}

\author[1,2]{Sina Feldmann}
\ead{sina_feldmann@brown.edu}
\author[1]{Juliane Adrian\corref{cor1}} 
\ead{j.adrian@fz-juelich.de}
\author[1]{Maik Boltes}

\cortext[cor1]{Corresponding author}
\address[1]{Institute for Advanced Simulation 7: Civil Safety Research, Forschungszentrum J\"ulich, J\"ulich, Germany}
\address[2]{Faculty of Architecture and Civil Engineering, University of Wuppertal, Wuppertal, Germany}

\begin{document}
\begin{frontmatter}

\begin{abstract}
Impulse propagation in crowds is a phenomenon that is crucial for understanding collective dynamics, but has been scarcely addressed so far.
Therefore, we have carried out experiments in which persons standing in a crowd are pushed forward in a controlled manner.
Variations of experimental parameters include (i) the intensity of the push, (ii) the initial inter-person distance, (iii) the preparedness of participants and (iv) the crowd formation.
Our analysis links the intensity of an impulse recorded by a pressure sensor with individual movements of participants based on head trajectories recorded by overhead cameras and 3D motion capturing data.

The propagation distance as well as the propagation speed of the external impact depends mainly on the intensity of the impulse, whereas no significant effect regarding the preparedness of participants could be found.
Especially the propagation speed is influenced by the initial inter-person distance.
From the comparison between two methods that detect the time of motion due to the impulse, a more sensitive result is obtained when the velocity of three landmarks of the human body is taken into account and not only the forward displacement of the center of mass.
Furthermore, the more intertwined participants are in relation to each other, the more the impulse is distributed to the sides. As a result, more people are affected, however with smaller individual displacements.
\end{abstract}

\end{frontmatter}

\section{Introduction}
\label{sec:intro}

Pedestrian models are valuable tools for understanding and managing crowd movement.
Due to their flexibility and cost-effectiveness, they can be employed for a variety of applications \cite{meyers_modelling_2018}.
For example, simulations enable safety to be assessed in crowded environments and help to identify potential bottlenecks, evacuation routes and areas prone to congestion \cite{korbmacher_review_2022}. 
This is particularly important for emergency scenarios and large events, where many people come together. There, various collective dynamics can occur \cite{meyers_empirical_2018}.
The investigation of these collective movements in crowds has become increasingly important in the context of safety, and computer-based approaches are applied more and more.

Despite recent developments in models, such as the simulation of a shock wave \cite{van_toll_extreme-density_2020}, the integration of forces in crowds \cite{song_experiment_2019} or the propagation of a disturbance \cite{li_disturbance_2023}, there are still limitations to overcome.
Pedestrian models are often based on simplifications and assumptions about human behaviour that cannot fully capture the complexity of real interactions.
For instance, people are often represented as circles or ellipses and the third dimension of movements is neglected, greatly simplifying the interactions between them. 
Even though there are models that attempt to simulate contact forces \cite{kim_velocity-based_2015, van_toll_sph_2021, wang_modelling_2023}, the propagation of impulses is particularly lacking.

One way to mitigate this problem is the "domino model" \cite{wang_modeling_2019}.
However, people do not necessarily behave like dominoes and different movements such as steps are not considered.
In tackling this complexity, understanding the advantages and disadvantages of established models is needed to better capture the complicated dynamics in dense crowds.

In order to validate and refine existing or new approaches, experimental data for comparison is crucial.
Preliminary small-scale studies investigate collision dynamics \cite{wang_study_2018}, contact forces \cite{li_experimental_2020} and propagation of impulses \cite{feldmann_forward_2023}, but are limited to 5 people.
This makes it difficult to transfer these results to large crowds.
Therefore, in the presented article, we extend the number of participants in a standing crowd that is perturbed by an external impulse up to 36.  
The impulse is here defined as the transmission of a force over a certain period of time.
Based on recorded trajectories, pressure data and 3D motions of the participants, we propose a propagation speed, propagation distance as well as a propagation angle dependent on the initial inter-person distance.
As discussed below, these quantities can directly be used for the validation of models.

\section{Methods}
\label{sec:1}

The propagation of an external impulse has already been studied in a small scale study for a row of five people in \cite{feldmann_forward_2023}.
We were able to draw first conclusions about how people regain their balance and interact with other people while transferring the impulse forward along a queue.
In order to better assess how this analysis can be applied to larger groups, we extended the five-row experiments.
Here, we will focus on the propagation of external impulses along a row of 20 people, as well as groups of up to 36 persons standing in different formations.
This enables us to draw statistically more accurate conclusions from a larger sample size and expand the analysis to two dimensions.
We vary different formations to achieve a more realistic representation of crowds and to investigate the influence of the positioning of people on the propagation of impulses. 

\subsection{Experiments}

The experiments were conducted in Wuppertal in May 2022 as part of the EU-funded project CrowdDNA \cite{CrowdDNAProject2022}. 
A detailed description of these experiments is summarised in the deliverable report D1.2 of the project \cite{D1.2_CrowdDNAProject2022}.
Furthermore, the description with a complete list of conducted trials as well as the collected datasets can be accessed from the Pedestrian Dynamics Data Archive \cite{DataArchive_impusle_2024}.

\begin{figure}[t]
\centering
\includegraphics[width=\textwidth]{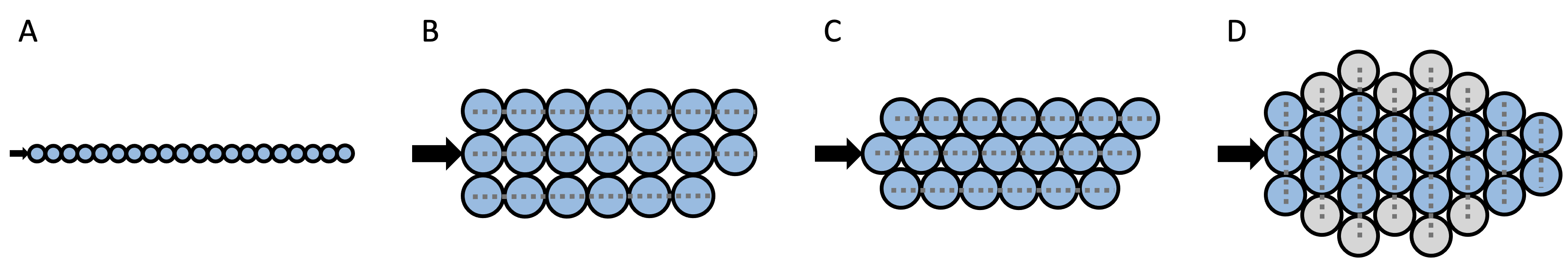}
\caption{Participants stand in various group formations before the push. (A) 20 people line up in a queue. 
(B) With their shoulders next to each other, participants stand in either three or five rows. 
(C) Either three or five rows standing next to each other are shifted to the front so that shoulders are positioned in the gap between two people.
(D) The people stand staggered so that one person has the shoulders of two people in front of them. The main group of participants wear a 3D motion capture (MoCap) suit (blue circles) and in some trials the sides are filled up with participants without a MoCap suit (grey circles). The external impulse from the punching bag is indicated as black arrow. Sketches of all investigated crowd formations can be found at \cite{DataArchive_impusle_2024}.}
\label{fig3:formations}
\end{figure}

The experiments can be divided into four blocks (block 1-4) in which different volunteers between the ages of 19 and 36 were recruited.
The experimental area was covered with judo mats to ensure a level of safety.
All participants were facing in the same direction away from the punching bag. They were positioned either as close as possible or at elbow distance in the prescribed group formation.
The tested formations (see \fref{fig3:formations}) were a row of 20 people (formation A), several rows (either three or five) standing shoulder to shoulder (formation B), the rows standing shifted to the front (formation C) or standing staggered, that each person had the shoulder of 2 people in front of them (formation D).
In total, there are 84 trials for formation A with 20 participants and 48 trials for each of the formations B, C and D were performed with a varying number $N$ of participants ($N \in [20, 36]$).
To achieve better comparability, the external impulses were carried out in the same way as in the five-row experiments.
The same experimenter manually pushed the punching bag, that was hanging horizontally from the ceiling, towards the back of the rearmost participant in the middle row at shoulder height. 
The pushing intensity was again alternated from weak to medium to strong and after these three pushes, the participants were repositioned in a controlled manner. 
In order to manipulate the participant's preparedness, the impulse occurred either directly announced (prepared condition) or after a random waiting time in which the participants had to recite the alphabet backwards (unprepared condition).
In these experiments, no instructions were given regarding the initial arm positions.

\subsection{Data sets}

\begin{figure}[t]
\centering
\includegraphics[width=\textwidth]{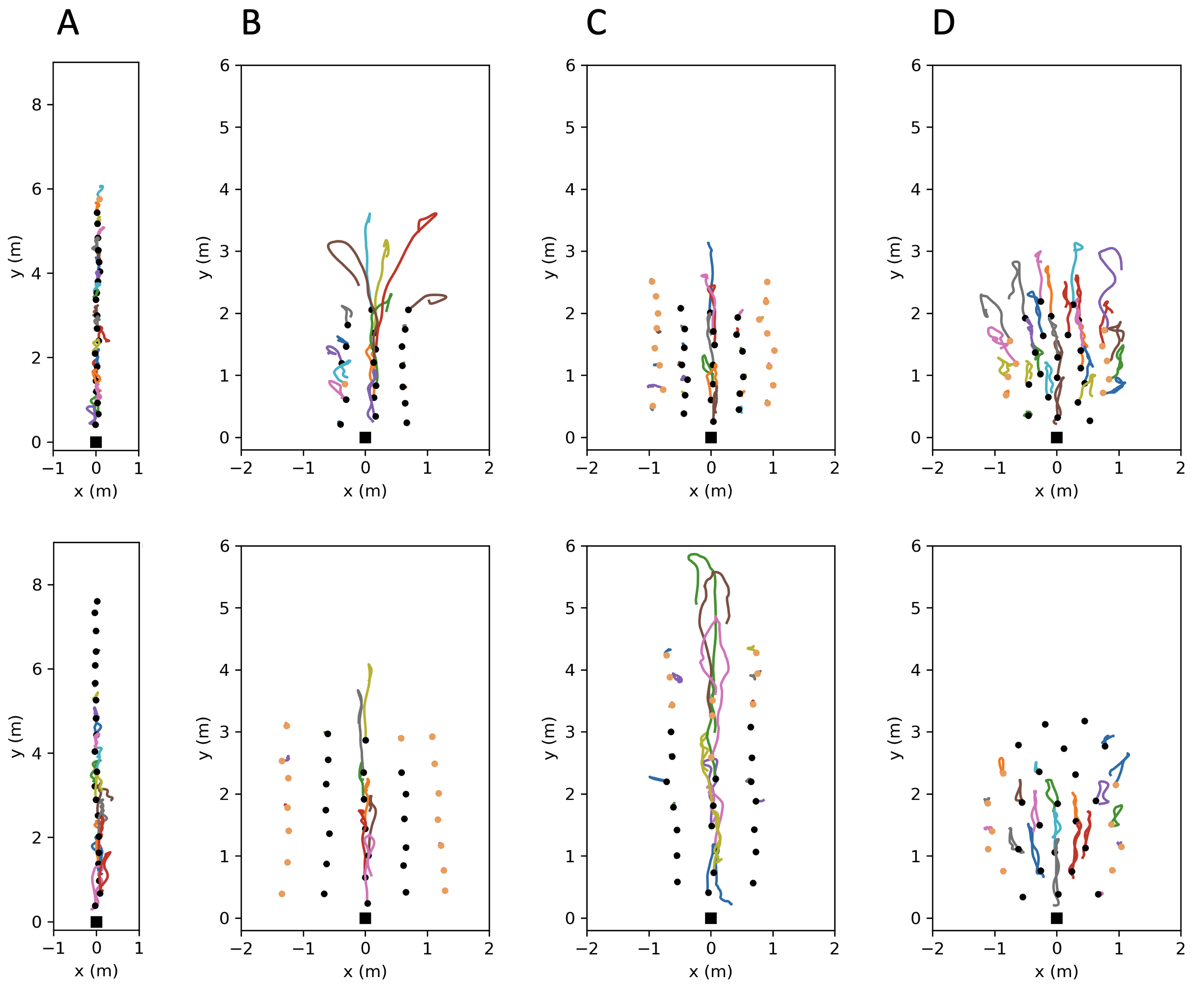}
\caption{Exemplary representation of collected trajectories for formation A, B, C and D with strong pushes. The upper plots show trials at no distance and the bottom plots at elbow distance. The initial positions are shown as black dots for participants in a MoCap suit, as light-orange dots for participants without MoCap suit and a black square for the punching bag. The middle row was pushed along the positive $y$-direction.}
\label{fig3:trajectories}
\end{figure}

Several video cameras captured the experiments, with the overhead cameras being particularly important as they allow the orange hats of each participant to be tracked with PeTrack, collecting individual head trajectories \cite{BOLTES2013127, boltes_maik_2021_5126562}.
\fref{fig3:trajectories} shows examples of collected trajectories for each formation at none (top) and elbow distance (bottom).
3D motion data of the individual limbs for 20 people were recorded using Xsens motion capture (MoCap) suits \cite{schepers_xsens_2018} that use 17 inertial measurement units (IMU). 
The advantage of these sensors is that they are based on relative measurements and do not rely on a line of sight, e.g. to cameras, which means that occlusions in groups have no influence on the quality of the MoCap data.
A disadvantage of these sensors is that the data is only displayed for one person at a time, which can be solved by mapping the 3D head data onto the camera trajectories \cite{boltes_hybrid_2021}. 
As a result, the 3D data is combined into a common coordinate system of the experimental area and therefore all participants are placed and oriented with respect to each other (see \fref{fig3:MoCap}, as well as Fig. S1 and Fig. S2 in the Supplementary Information).
By attaching an Xsensor \cite{XsensorLX210:50.50.05} pressure sensor to the front of the punching bag, the intensity of the impulses was determined.
\\

\begin{figure}[t]
\centering
\includegraphics[width=0.9\textwidth]{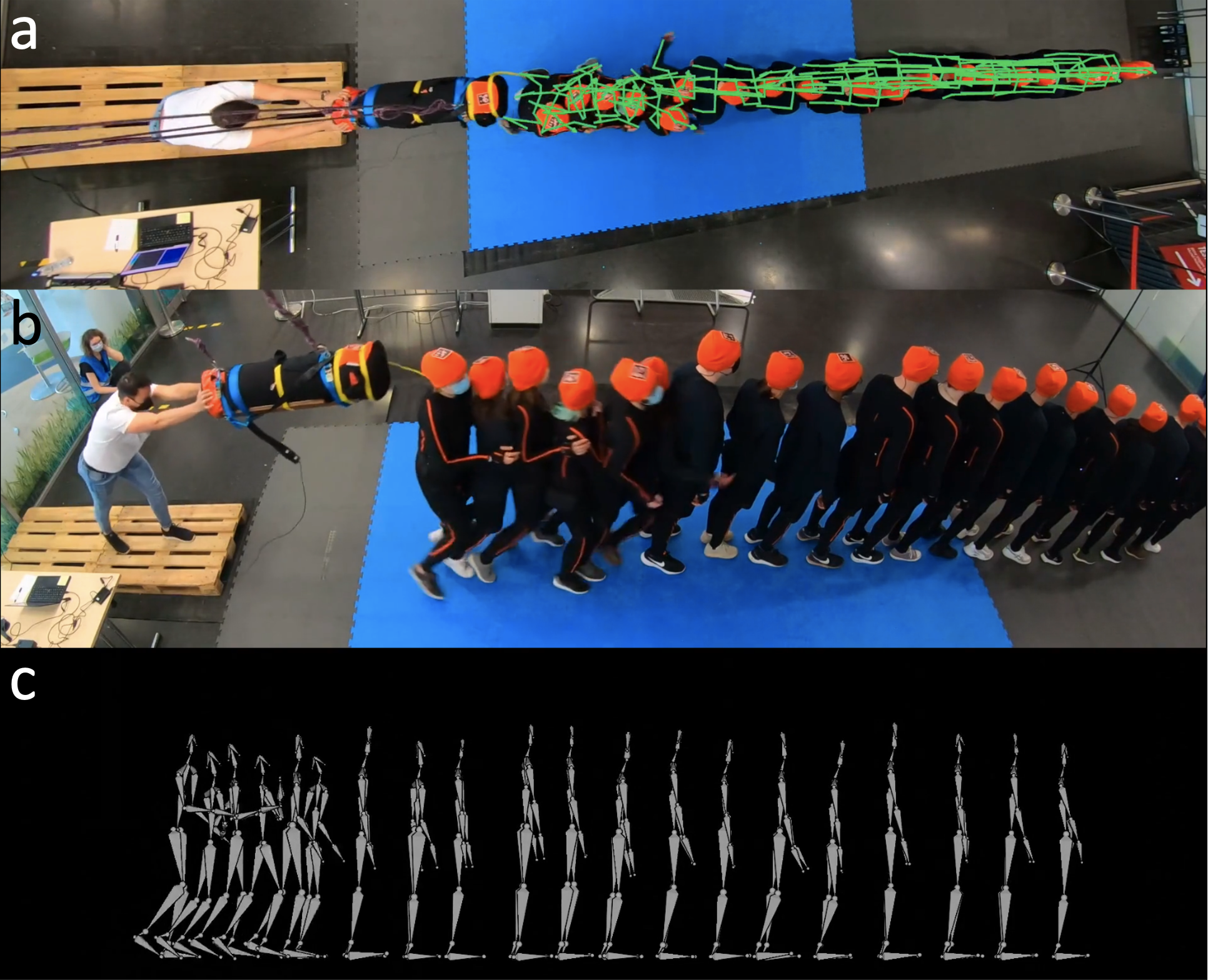}
\caption{Example of a trial after 2\,s of the push, during which 20 participants lined up in formation A at elbow distance and the rearmost person was pushed forward by the punching bag. (a) Snapshot of the overhead camera used to extract head trajectories from the orange hats. The combined 3D MoCap data with exact positioning in the experimental area are shown as green stick figures. (b) Snapshot of a side-view camera enabling a qualitative analysis. (c) Visualisation in Blender of the combined MoCap data of all 20 participants from the side. Examples of formation B, C and D can be found in the Supplementary Information (see Fig. S1 and S2).}
\label{fig3:MoCap}
\end{figure}

Since the external impulse affects the whole body and participants can react with different complex movements, it is important to consider more than just head movements. 
The center of mass (CoM) is a suitable measure for investigating impulse propagation as it includes a wide range of movements and can be determined from the 3D MoCap data.
3D motion data were collected for the main group, i.e. 20 persons in the middle of each formation. Due to technical issues, a 3D data set for one person is missing in half of the trials.
To account for boundary effects, the formations were filled up at lateral or frontal positions with participants for whom only head trajectories were recorded.
In our analysis, we solely consider the combined 3D MoCap data.
The experimental data including videos, head trajectories, 3D MoCap data as well as pressure data can be freely accessed from the Pedestrian Dynamics Data Archive of the Research Centre J\"ulich \cite{DataArchive_impusle_2024}.

\subsection{Analysis}
\label{sec:analysis}

In our analysis, we will compare the forward propagation of the external impulse of the experiments shown in \fref{fig3:formations}A for 20 people in a row with the results of the five-row experiments \cite{feldmann_forward_2023}, contrast the methods for motion detection from \cite{feldmann_forward_2023} and \cite{feldmann_temporal_2024} and extend the analysis to the side for formations B, C and D.

\begin{figure}[htbp]
\centering
\includegraphics[width=\textwidth]{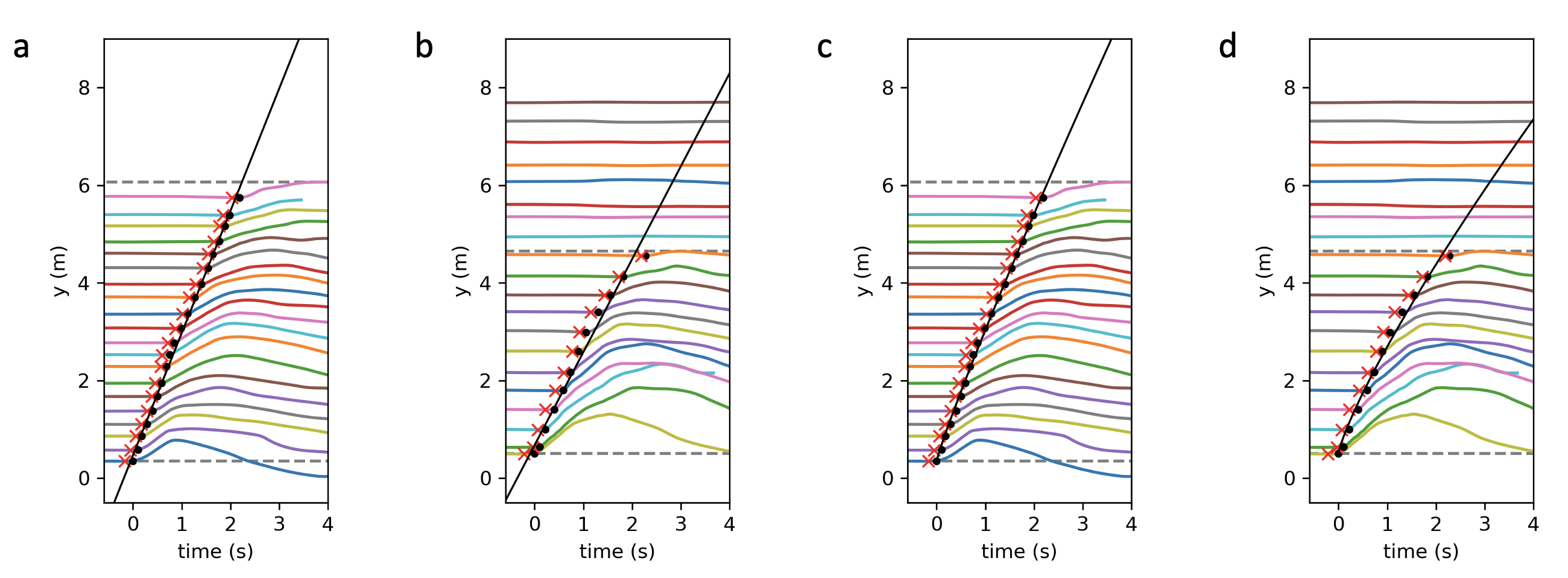}
\caption{Changes in the $y$-position of the CoM over time indicate a forward motion of the participants when pushed from behind. The participants stand in formation A at no distance (a) or elbow distance (b). Using the elbow method, the start of motion (black dots) for each person is determined if they are affected by the push. The gradient of the linear regression through these points results in a forward propagation speed. The start times detected by the velocity-method (\eref{eq:velo2}) are shown as red crosses. (c) and (d) show the same experimental trials as (a) and (b) with the difference that a power function is fitted to the elbow points. The coefficient b of the fitted functions indicates that (c) a linear fit is justified or (d) the impulse is dampened along the row.}
\label{fig3:y-t-plot}
\end{figure}

For a better comparison, the same analytical methods as in \cite{feldmann_forward_2023} are applied to investigate the propagation speed and the propagation distance. 
This means that only the forward trajectories of the CoM, i.e. the $y$-direction, are examined over time as shown in \fref{fig3:y-t-plot}.
Based on this, the start times as well as positions of the persons affected by the external impulse are detected with elbow points (marked as black dots).
The elbow point corresponds to the point at which the y-t curve bends visibly, and is often used in mathematics for optimization or cluster analysis.
In order to find this point, the y-t-curves between the initial value $y_i^{\text{ ini}} = y_i(t_i^{\text{ ini}})$ and the maximum forward displacement $y_i^{\text{ end}} = y_i(t_i^{\text{ end}}) = \max[y_i (t)]$ are considered.
Hereby, $i$ indicates the number of the person in the row ($i \in [1,20]$) and the initial time is set as $t_i^{\text{ ini}} = 1.5\,$s before the push $\forall i$.
The elbow point is the point of the curve that has the greatest perpendicular distance to the line $\mathbf{\tilde{g}}_i$.
The line $\mathbf{\tilde{g}}_i$ passes trough the two points $\mathbf{\tilde{a}} = (y_i^{\text{ ini}}, t_i^{\text{ ini}})$ and $ \mathbf{\tilde{b}} = (y_i^{\text{ end}}, t_i^{\text{ end}})$ and is defined as $\mathbf{\tilde{g}}_i = \mathbf{\tilde{a}} + k \cdot \mathbf{\tilde{h}} $ with normalized direction vector $\mathbf{\tilde{h}} = \frac{(\mathbf{\tilde{b}} - \mathbf{\tilde{a}})}{ \| (\mathbf{\tilde{b}} - \mathbf{\tilde{a}})\|}$ along the line and $k \in \mathbb{R}$.  
The elbow point is only calculated, when person $i$ has moved forward more than $0.065\,\text{m}$.
\\

The gradient of the resulting regression line (black line) through these elbow points results in the propagation speed.
First, a linear fit is selected to allow a closer comparison with \cite{feldmann_forward_2023}, as the elbow points were always aligned linearly for the five-row experiments.
However, an impulse could also be dampened or intensified along the row. 
In order to test all three hypotheses at once, a power function is fitted in a subsequent analysis (see \fref{fig3:y-t-plot}\,c and \fref{fig3:y-t-plot}\,d and Section \ref{chap:propagation_speed})
\\

The propagation distance of the push is calculated as the distance between the starting position of the person standing at the punching bag and the furthest forward position of $j$, which is the last person moving.
This is represented as grey dashed lines in \fref{fig3:y-t-plot}.

\begin{equation}  
d_{\text{push}} = \max[y_j(t_j) - y_1(0)]
\label{eq:d_push}
\end{equation} 

Using the data collected by the pressure sensor, the impulse can be determined by integrating the pressure $P$ over the sensor area $A$ as well as over time $t$.
\begin{equation}  
J_{\text{bag}} = \int_t \int_A P \, \text{d}A \, \text{d}t
\label{eq:impulse}
\end{equation} 
\\

In a second step, the elbow-method for detecting the start of motion as result of the external impulse is compared with another method that considers the velocity of three different landmarks of the human skeleton simultaneously. 
This velocity-method is introduced in \cite{feldmann_temporal_2024}, which is an extended investigation of the five-row experiments.
Hereby, individual 3D reactions of participants are separated into temporal phases and the velocity defines the start of the first phase of motion.
With the velocity method, a motion as response to the external impulse is detected if the forward acceleration of the CoM exceeds $0.3\, \frac{\text{m}}{\text{s}^2}$ and if shortly afterwards the CoM, the lowest cervical vertebra C7 and the origin of the hip move forward at a velocity faster than $0.05\, \frac{\text{m}}{\text{s}} $.
The actual start time of the motion is then set to the time when the forward acceleration of the CoM equals $0.15\, \frac{\text{m}}{\text{s}^2}$.

Forward velocities $v_i^p(t)$ and forward accelerations $a_i^p(t)$ for person $i$ can be calculated from the $y$-position of individual body parts as follows:

\begin{equation} 
v_i^p(t) = \left(\frac{y_i^p(t+\Delta t) - y_i^p(t-\Delta t)}{2 \cdot \Delta t} \right)
\label{eq:v}
\end{equation}

\begin{equation} 
a_i^	p(t) = \left( \frac{v_i^p(t+\Delta t) - v_i^p(t-\Delta t)}{2 \cdot \Delta t} \right)
\label{eq:a}
\end{equation}

Hereby, $\Delta t = 0.05 \,\text{s}$ was used and $p$ corresponds to an anatomical landmark of the human body.
With the velocity method, a start of motion due to a push is only calculated when the following condition is at least once true:

\begin{equation*} 
\scalebox{0.9}{$
\exists \; t_1 < t_2:\, a_i^{\text{CoM}}(t_1) > 0.3\, \frac{\text{m}}{\text{s}^2} \land  
   v_i^{\text{CoM}}(t_2)> 0.05\, \frac{\text{m}}{\text{s}} \land 
   v_i^{\text{C7}}(t_2) > 0.05\, \frac{\text{m}}{\text{s}}\land 
   v_i^{\text{HIP}}(t_2) > 0.05\, \frac{\text{m}}{\text{s}}
  \label{eq:velo1}
  $}
\end{equation*}

The actual start time is then calculated with:
\begin{equation} 
t_i^{\text{ VEL}} = \max[t_0], \quad \text{if } a_i^{\text{CoM}}(t_0) > 0.15\, \frac{\text{m}}{\text{s}^2}, \quad \text{for } t_0 < \min[t_2]
\label{eq:velo2}
\end{equation}

where $t_0, t_1, t_2 \in [0\,\text{s}, 5\,\text{s}]$ after the push. For further details of the definition, please refer to \cite{feldmann_temporal_2024}.
\\

In order to investigate the impact of the impulse to the lateral standing participants, data from formations B, C and D are also included. 
The start positions of the participants $\mathbf{\tilde{r}}_i^{\text{CoM}}(0)$ in relation to the punching bag $\mathbf{\tilde{r}}_i^{\text{BAG}}(0)$ in the $xy$-plane are taken into account.
For $\mathbf{\tilde{r}}_i^{\text{CoM}}, \mathbf{\tilde{r}}_i^{\text{BAG}}, \mathbf{l}_i, \mathbf{e}_y \in \mathbb{R}^2 $, a distance $l_i = \|\mathbf{l}_i \|$ as well as an angle $\alpha_i $ to the $y$-direction are determined:
\begin{equation} 
\alpha_i = \text{sgn}(\mathbf{l}_i ) \cdot \arccos \left( \frac{ \mathbf{l}_i \cdot  \mathbf{e}_y}{\|\mathbf{l}_i \|} \right)
\label{eq:angle}
\end{equation}

with $ \mathbf{l}_i = \mathbf{\tilde{r}}_i^{\text{CoM}}(0) - \mathbf{\tilde{r}}_i^{\text{BAG}}(0)$, $ \mathbf{e}_y = \binom{0}{1}$ and 
$\text{sgn}(\mathbf{l}_i ) = 
\begin{cases}
    +1, & \text{if } x_i > 0\\
    -1, & \text{otherwise}
\end{cases}
$.\\

The displacement of each participant is calculated as the distance between the starting position $\mathbf{\tilde{r}}_i^{\text{CoM}} (0)$ and points on the trajectory $\mathbf{\tilde{r}}_i^{\text{CoM}}(t)$ in the $xy$-plane.
The maximum displacement $\Delta s_i^{\text{ max}} $ is obtained accordingly:

\begin{equation}
\Delta s_i^{\text{ max}}  =  \max[  \| \mathbf{\tilde{r}}_i^{\text{CoM}}(t) - \mathbf{\tilde{r}}_i^{\text{CoM}} (0) \| ]
\label{eq:displacement}
\end{equation}
\\

The analysis was executed in Python and the \textit{ezc3d} library \cite{michaud_ezc3d_2021} was used for the 3D MoCap data.
Statistical tests were carried out in R version 4.3.2 and the outputs can be found in the Supplementary Information (see Tables S1 - S6).

 \section{Results }
 
\subsection{Comparison to five-row experiments}

\subsubsection{Preparation}

\begin{figure}[htbp]
\centering
\includegraphics[width=0.8\textwidth]{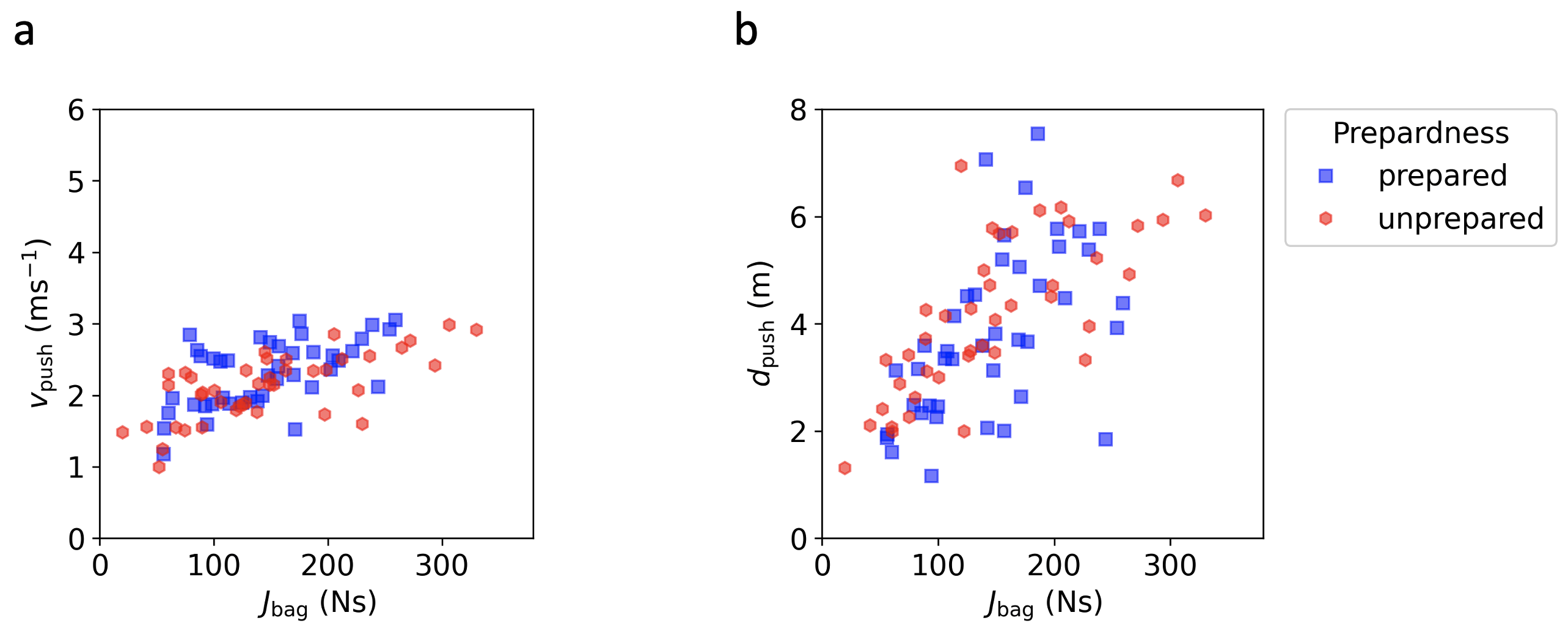}
\caption{(a) Propagation speed and (b) propagation distance of the push through the row of 20 people (formation A) as functions of the impulse measured at the punching bag. The propagation of the external impulse is not significantly influenced by the preparedness of the participants.}
\label{fig3:preparation}
\end{figure}

First, the influence of the preparation level of the participants (prepared or unprepared) will be examined as this aspect was not considered in the five-row experiments.
 \fref{fig3:preparation} indicates that the preparedness of the participants has neither a significant effect on the propagation speed $v_{\text{push}}$ nor the propagation distance $d_{\text{push}}$ in relation to the impulse $J_{\text{bag}}$ at the punching bag when the participants stand in a long queue of 20 people. 
This aspect is therefore neglected in the subsequent analysis and no distinction is made between prepared and unprepared trials.

\subsubsection{Propagation speed}
\label{chap:propagation_speed}

\begin{figure}[htbp]
\centering
\includegraphics[width=0.9\textwidth]{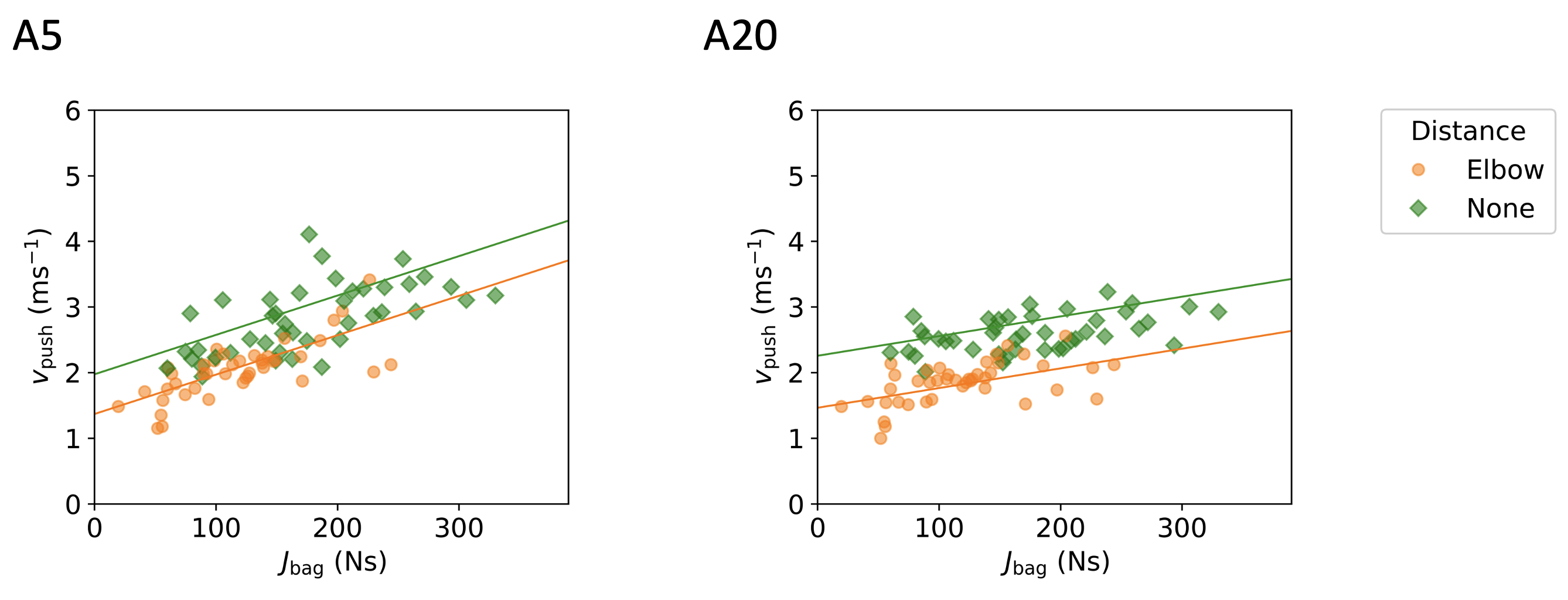}
\caption{Propagation speed as a function of the measured impulse for the formations A5 and A20. There is a significant difference between the initial inter-person distances none and elbow. A20 differs significantly from A5, which indicates that an impulse can be absorbed along a long row.}
\label{fig3:propagation_speed}
\end{figure}

\cite{feldmann_forward_2023} come to the conclusion that the propagation speed is linearly correlated with the impulse of the push, but does not depend on the initial inter-person distance.
To investigate the propagation of the external impulse in more detail, we have carried out experiments on a larger scale. 
For a comparison between the two experiments, trials from the five-row experiments with a free initial arm posture at none and elbow distance are included.
In the up-scaled experiments described here, all trials of formation A are evaluated and the propagation speeds through the first five persons of the long rows (A5) are calculated. 
A linear correlation between the propagation speed and the impulse is confirmed (see \fref{fig3:propagation_speed}).
However, a moderation analysis reveals a significant difference between none and elbow distance without interaction, that was previously undetected (see Table S1 in the SI).
It is important to note, that the five-row experiments are comparable to A5 at no distance, but significantly differ at elbow distance (see Tables S5 and S6 in the SI).

The moderation analysis results in the following two equations for the propagation speed for formation A5:

$$\text{A5:}\quad v_{\text{push, none}} = 0.006 \, \frac{\text{m}}{\text{Ns}^2} \cdot J_{\text{bag}} + 1.974 \, \frac{\text{m}}{\text{s}}$$
$$\text{A5:}\quad v_{\text{push, elbow}} = 0.006 \, \frac{\text{m}}{\text{Ns}^2} \cdot J_{\text{bag}} + 1.369 \, \frac{\text{m}}{\text{s}}$$

These results differ significantly from the equations found for the five-row experiments \cite{feldmann_forward_2023}.
In particular, the gradients across experiments vary considerably.
Furthermore, it can be noted that the measured impulses are smaller in the five-row experiments compared to the up-scaled experiments.
\\

Now, the propagation speeds are calculated for all persons in the long rows (A20) and compared to the first 5 persons in the long rows (A5).
A moderation analysis also reveals a significant effect of the initial inter-person distance on the propagation speed for A20 (see S2 in SI).
In addition, there are significantly lower propagation speeds at higher impulses for the long rows compared to A5 (see S5 and S6 in the SI).
This results in the two equations for the propagation speed for formation A20, as follows:
$$\text{A20:}\quad v_{\text{push, none}} = 0.003 \, \frac{\text{m}}{\text{Ns}^2} \cdot J_{\text{bag}} + 2.256 \, \frac{\text{m}}{\text{s}}$$
$$\text{A20:}\quad v_{\text{push, elbow}} = 0.003 \, \frac{\text{m}}{\text{Ns}^2} \cdot J_{\text{bag}} + 1.464 \, \frac{\text{m}}{\text{s}}$$
\\

\begin{figure}[htbp]
\centering
\includegraphics[width=0.6\textwidth]{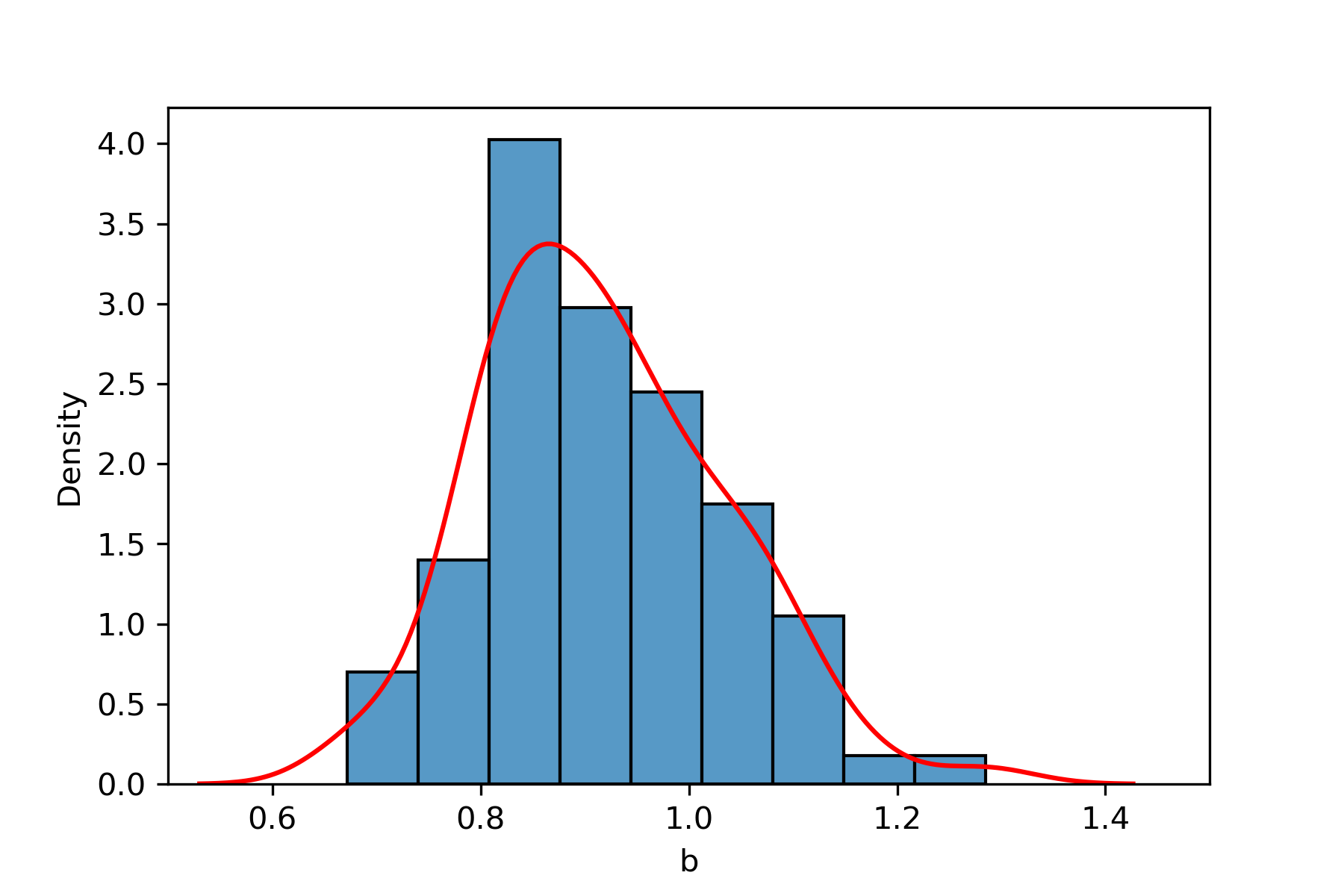}
\caption{Distribution of the coefficient $b$ obtained from a power function fit of the elbow points for A trials. The maximum is less than 1, which suggests an absorption effect along the row of 20 participants.}
\label{fig3:hist_b}
\end{figure}

This could indicate that the impulses are being dampened along the rows of 20 people.
To investigate this, we fitted a power function $y(t) = a \cdot t^b + c\,$ to the elbow points of A20 instead of the linear fit.
Two examples of the power function fit can be found in \fref{fig3:y-t-plot}\,c and \fref{fig3:y-t-plot}\,d.
The coefficient $b$ is of particular interest here, as it specifies whether a linear fit is justified ($b \approx 1$), whether the impulse is absorbed ($b < 1$) or whether the impulse is intensified ($b>1$). 
\fref{fig3:hist_b} shows the distribution as well as the kernel density estimate of $b$ for all A trials. 
It can be seen that all three cases occur.
However, the maximum of the distribution lies below 1, indicating that on average pushes are dampened when they propagate through the long row.

\subsubsection{Propagation distance}

\begin{figure}[htbp]
\centering
\includegraphics[width=0.6\textwidth]{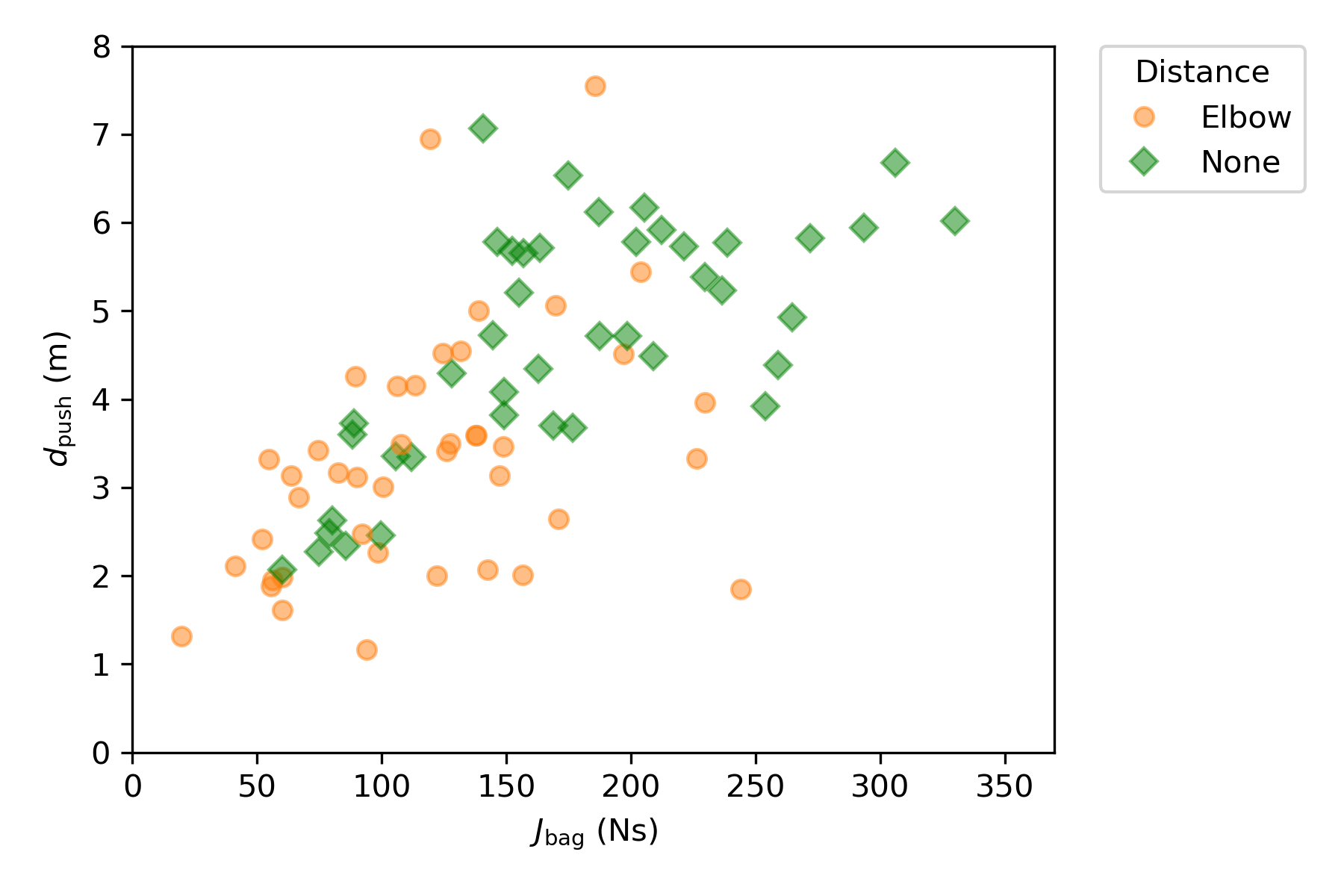}
\caption{Propagation distance for the long rows A20 increases with stronger impulses. There is no difference between none and elbow distance.}
\label{fig3:propagation_distance}
\end{figure}

\fref{fig3:propagation_distance} shows the propagation distance (\eref{eq:d_push}) as a function of the measured impulse (\eref{eq:impulse}).
A linear correlation between the propagation distance and the impulse is further supported by this analysis and no effect of the initial inter-person distance on the propagation distance can be found.

These results are difficult to compare with the five-row experiments, as the push propagates through the entire row of 5 people starting at $J = 110$\,N and therefore, the propagation distance reaches a maximum of approx. 3\,m.
However, the distance is also bounded in the large experiments, as the impulse sometimes passes through the entire row of 20 people (see \fref{fig3:y-t-plot}\,a).
This indicates, that the distance should be investigated using even more people.

\subsection{Comparison of the two detection methods}

We proposed two different methods (see section \sref{sec:analysis}) to detect whether a participant is affected by the external impulse and to determine the time the motion starts.
In \fref{fig3:y-t-plot}, the elbow-method is shown as black dots and the velocity method derived from \eref{eq:velo2} as red crosses.
It can be seen that the two methods are well comparable with each other, whereby the velocity method sets a slightly earlier point in time.
A qualitative comparison of both methods for all A trials with the recorded videos shows that the velocity method is more sensitive than the elbow method. 
The number of falsely detected persons are listed in \tref{tab:method}.

\begin{table}[htbp]
\centering
    \caption{Comparison of the elbow method as well as the velocity method with the recorded videos for A trials.}
    \label{tab:method}
    \begin{tabular}{l|cc|c}
    	\toprule
        method & false -positive &  false-negative & total \\
         \midrule
 velocity & 29 & 13 & 1644 \\
elbow & 8 & 90 & 1680 \\
\bottomrule
    \end{tabular}
\end{table}

\subsection{Extension to side}

The formations B, C and D are included to extend the analysis to the side.
First, we want to investigate the effect of lateral standing rows on the propagation speed of the middle rows for formations B and C.
Therefore, the same analysis as discussed in Section \sref{chap:propagation_speed} is carried out for the middle row, which stands directly in front of the punching bag. 
The moderation analyses show no significant difference between the initial inter-person distances for both formations (see Tables S3 and S4 in the SI).
For the propagation speed, the equations 
$v_{\text{push}} = 0.005 \, \frac{\text{m}}{\text{Ns}^2} \cdot J_{\text{bag}} + 1.658 \, \frac{\text{m}}{\text{s}}$ and $v_{\text{push}} = 0.004 \, \frac{\text{m}}{\text{Ns}^2} \cdot J_{\text{bag}} + 1.666\, \frac{\text{m}}{\text{s}}$ are obtained for formations B and C respectively. 
\\

\begin{figure}[htbp]
\centering
\includegraphics[width=\textwidth]{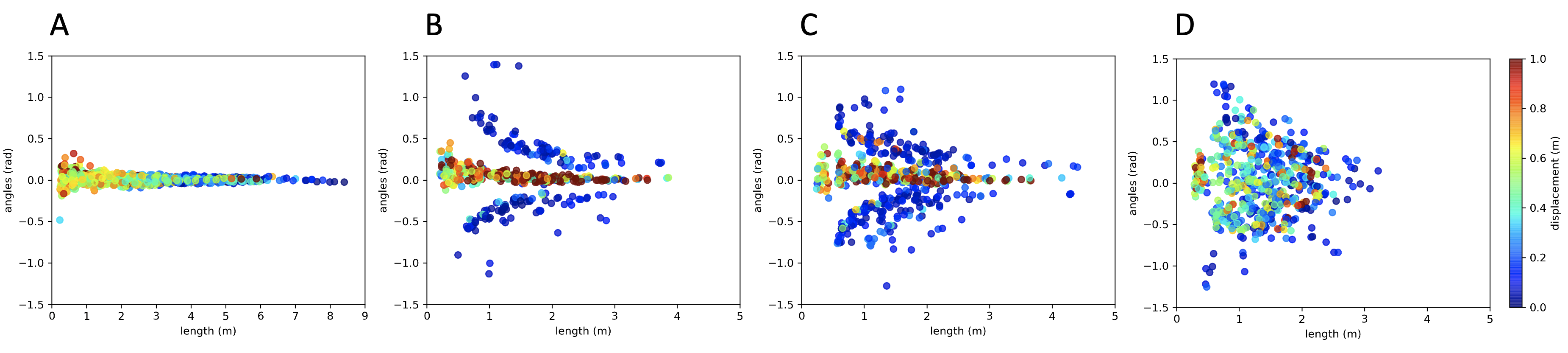}
\caption{Heat map of the individual displacement of the initial position for formations A, B, C and D. Angles $\alpha_i$ are shown as function of the distance $l_i$ from the punching bag to the initial positions of participants that are affected by the external impulse (\eref{eq:angle}). The color indicates the maximum displacement $\Delta s_i^{\text{ max}}$ of the participant (\eref{eq:displacement}).}
\label{fig3:propagation_side}
\end{figure}

In order to investigate the impact of the impulse to the lateral standing participants, the velocity method (\eref{eq:velo2}) is used to detect the start positions of participants that are affected by the push.
For these participants, the distance $l_i$ as well as the angle $\alpha_i$ to the start position of the punching bag are calculated with \eref{eq:angle}. 
The angle becomes negative if the person stands on the left of the punching bag and positive when standing on the right side.
\fref{fig3:propagation_side} shows a heat map of the determined angles as a function of distance for formations A, B, C and D.
The different colours of the points represent the maximum displacement $\Delta s_i^{\text{ max}}$ of the participant.
This is calculated as the distance between the starting position and the furthest point on the trajectory in the $xy$-plane (see \eref{eq:displacement}).

The more intertwined the participants are standing in relation to each other, the more lateral components occur.
The external impulse is distributed over a larger area and therefore propagates less to the front and more to the sides.
As a result, the individual displacements are smaller for participants standing in formation D compared to B or C.

\section{Discussion}

This article investigates the propagation of external impulses through standing crowds.
For this purpose, experiments were carried out that illustrate a simplified representation of real crowds.
First, only one row of people is analysed, which is then extended to multiple rows and a group formation.

\subsection{Limitations}
One limitation of this study is manual pushing and the pressure sensor.
The perceived strength of the pusher can differ from the measured values, so the categorisation is not clear.
Besides, the pressure sensor only records a normal component of the pushing force which does not fully correspond to reality. 
The experiments are also restricted in comparison to real life scenarios. 
Since we are mainly investigating rows here, this analysis could mainly be applied to queuing systems. 
In real dense crowds, people probably do not stand as uniformly as in these experiments. 
Therefore, it would be interesting to investigate what happens when people are not standing evenly distributed and what effect a gap in the crowd can have on the propagation of impulses.
Another aspect contrary to reality is that the impulses only come from one direction and therefore other pushing directions could be investigated in the future.
Additional factors, that could be considered for a further and more realistic analysis are uneven ground or the awareness level of participants.
\\

\subsection{Conclusion}
In this study, we find that the propagation speed of an impulse depends on the intensity of the push and the initial inter-person distance.
A significant difference regarding the preparation of participants on the propagation could not be found.
Here it is particularly interesting to note that the initial inter-person distance actually causes a significant difference in the propagation speed, which could not be shown in the preceding five-row experiments \cite{feldmann_forward_2023}. 
We assume that this could be explained by a small number of repetitions for the same conditions and the same participants.
\\

Furthermore, a row of five people is quite short, so that there may be boundary effects and certain phenomena remain unrecognized.
Along the row of 20 people, an absorption effect of the impulse is observed.
When using a power function fit to investigate the damping of the push through the row, one assumes that there is a constant coefficient $b$ for the complete range of the propagation. 
This might be improved to a position-dependent absorption coefficient since we observe a linear propagation speed for the first part of the row.
The damping only starts for the last few participants affected by the push, resulting in a J-shaped propagation speed curve.
One explanation for this would be that persons behave similar to the non-Newtonian mixture of starch and water in the well-known popular experiment: They react more stiff to stronger impacts whereas with decreasing impact strength, muscles have time to counteract and dissipate the external force.
This implies that the absorption coefficient is rather a function of the intensity than the position.
On the other hand, of course, the intensity itself depends on the distance from the initial impact.
These findings underline the fact that the domino model is not a good choice for modelling impulse propagation through crowds. 
In addition, it should be noted that people are not physical particles, but individuals with different characteristics (e.g. height, weight) who can use different strategies to regain balance. 
All of this may have an impact on the variables we are investigating.
\\

On the other hand, it could also be possible that an impulse is intensified along the row, which is indicated by a few cases when the coefficient $b > 1$. 
We assume that different reactions of participants can be the reason for this.
The exact circumstances of this, however, require further investigation.
In addition, the propagation speed is only defined for one row and should be extended to 2 dimensions in the future.
\\

A big advantage of this evaluation is that we use 3D data to draw a 2D illustration of the propagation.
Especially, the investigation on how strong the impulses affect people, depending on their position within a crowd.
This heat map can be particularly helpful for models that rely on 2D representations of people.
\\
\\
\\

\noindent \textbf{Acknowledgements}\\
This study was funded by the European Unions Horizon 2020 research and innovation program within the project CrowdDNA [grant number 899739].
We would like to thank Jernej {\v C}amernik, Thomas Chatagnon, Helena L\"ugering, Armin Seyfried and Anna Sieben who supported us in the conception, planning and realisation of the experiments. Furthermore, we are grateful to Alica Kandler for her help in setting up the experiments and curating the data.
\\

\noindent \textbf{Ethics Statementt} \\
The experiments were approved by the ethics board of the University of Wuppertal in April 2022 (Reference: \mbox{MS/AE 220330}).
\\

\noindent \textbf{Authors contribution}\\
Sina Feldmann: Planning and conducting experiments, curating and analysing data, writing original draft / 
Juliane Adrian: Planning and conducting experiments, curating data, supervision, review and editing / 
Maik Boltes: Planning and conducting experiments, supervision, review and editing. 
All authors agree to the publication of the manuscript
\\

\noindent \textbf{Data Availability}\\
The raw data from the experiments are freely accessible at the Pedestrian Dynamics Data Archive of the Research Centre J\"ulich,
\mbox{\doi{10.34735/ped.2022.6}}.
\\

\bibliographystyle{cdbibstyle} 
\bibliography{pedbib} 

\clearpage

\newpage

\renewcommand{\thetable}{S\arabic{table}}
\renewcommand{\thefigure}{S\arabic{figure}}
\renewcommand{\thesection}{S}
\renewcommand{\thesubsection}{S\arabic{subsection}}
\setcounter{section}{0}

\markboth{{Publication C Propagation of controlled frontward impulses}}{{Supplementary Information}}
\section{Supplementary Information}
\markboth{{Publication C Propagation of controlled frontward impulses}}{{Supplementary Information}}

 \setcounter{subsection}{0}
 \setcounter{table}{0}
  \setcounter{figure}{0}

\subsection{Experimental Data}
\label{appendix:experiment}

The experiments were recorded with several video cameras. 
The top-down videos are used to collect individual head trajectories of each participant with PeTrack and the sideview camera enable a qualitative analysis. 
The 3D motion of 20 participants recorded with MoCap suits from Xsens are combined with the trajectories to integrate them into a common coordinate system of the experimental area.

\begin{figure}[htbp]
\centering
\includegraphics[width=\textwidth]{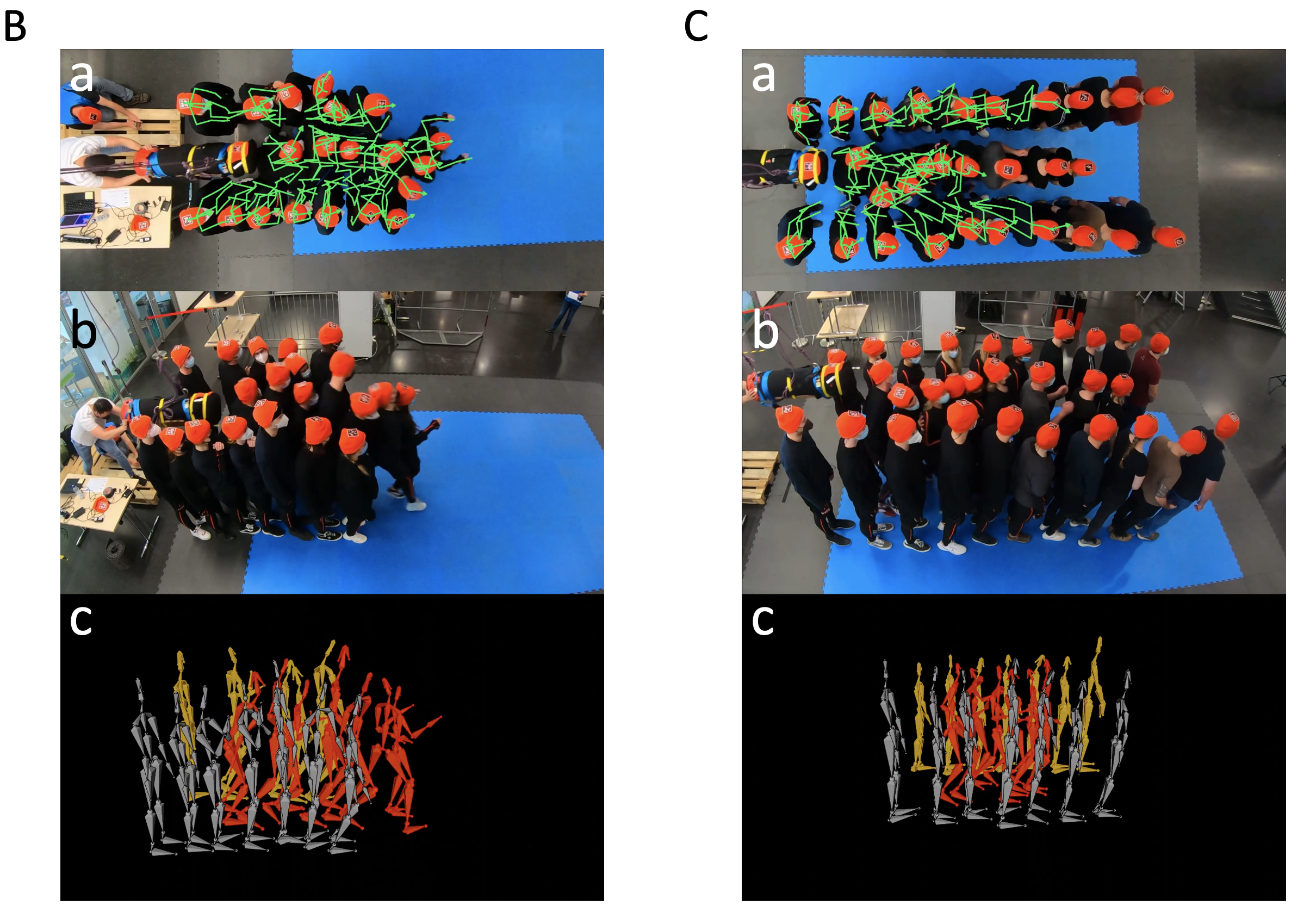}
\caption{Examples of formation B and C.
(a) Snapshot of the overhead camera with combined 3D MoCap data visualised as green stick figures.
(b) Snapshot of a side-view camera enabling a qualitative analysis. 
(c) Visualisation in Blender of the combined MoCap data of 20 participants from the side. The front row is coloured in grey, the middle row in red and the back row in yellow.}
\label{fig3:expB}
\end{figure}

\begin{figure}[htbp]
\centering
\includegraphics[width=0.5\textwidth]{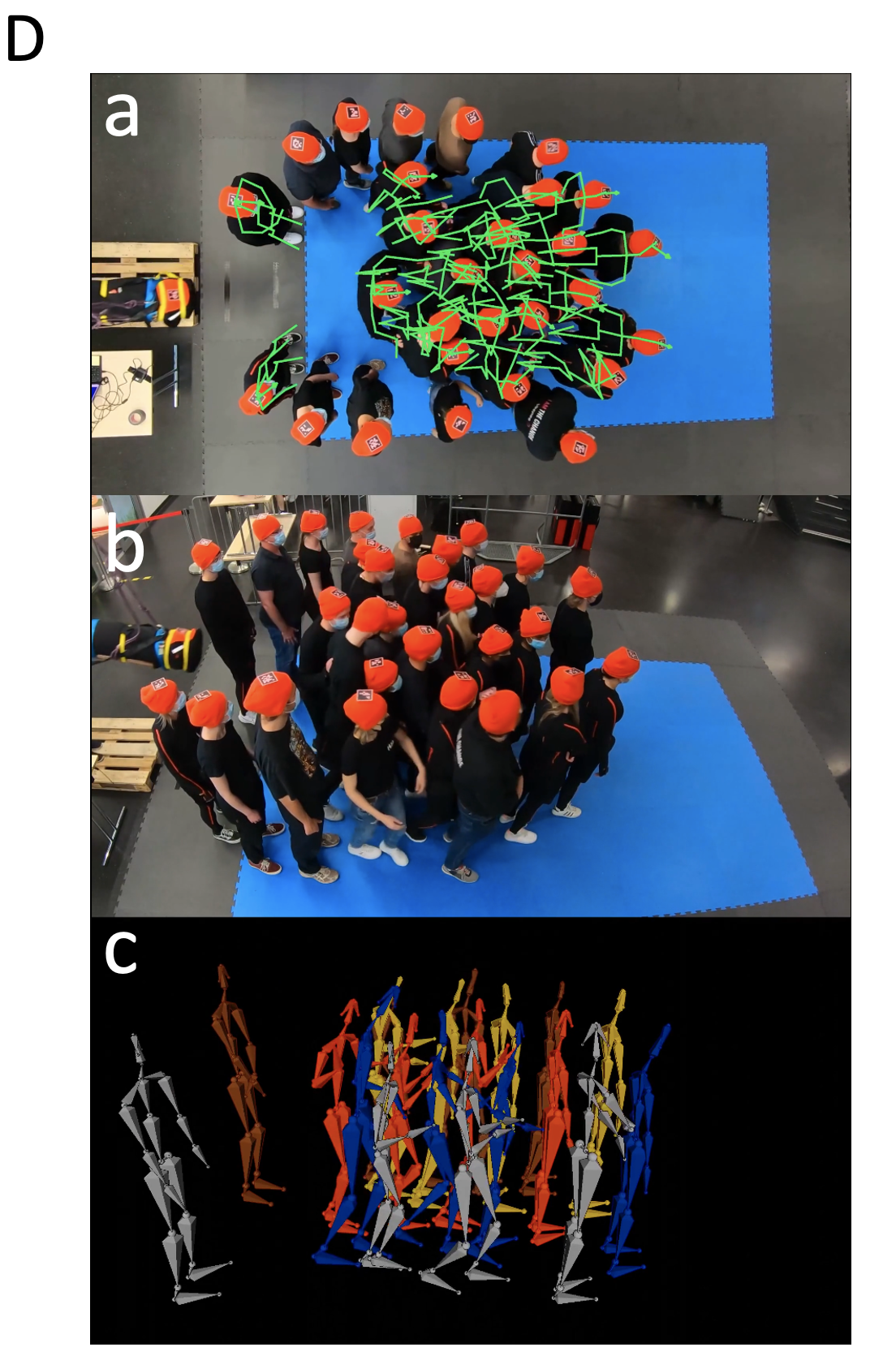}
\caption{Example of formation D
(a) Snapshot of the overhead camera with combined 3D MoCap data visualised as green stick figures.
(b) Snapshot of a side-view camera enabling a qualitative analysis. 
(c) Visualisation in Blender of the combined MoCap data of 20 participants from the side. The skeletons are coloured from front to back ($x$-position) as follows: grey, blue, red, yellow, brown.}
\label{fig3:expD}
\end{figure}

\newpage
\subsection{Moderation analysis}
\label{appendix:statistics}

All moderation analyses were conducted in R version 4.3.2. 
First, we compare the effect of initial inter-person distance for each formation A5, A20, B and C individually.
In doing so, only the middle row, which stands directly in front of the punching bag, is taken into account. 
The results are shown in Tables \ref{tab:statistics_a5}, \ref{tab:statistics_a20} , \ref{tab:statistics_b} and \ref{tab:statistics_c}  for the comparison of no and elbow distance.

\begin{table}[htbp]
\centering
    \caption{Results of the moderation analysis comparing no and elbow distance for A5.}
    \label{tab:statistics_a5}
    \begin{tabular}{lrrrll} 
    	\toprule
	\addlinespace
	 Residuals: &  \multicolumn{1}{l}{Min} &  \multicolumn{1}{l}{1Q} &  \multicolumn{1}{l}{Median} &  \multicolumn{1}{l}{3Q} &  \multicolumn{1}{l}{Max} \\
	 & \multicolumn{1}{l}{-0.78745 } & \multicolumn{1}{l}{-0.21077} & \multicolumn{1}{l}{-0.01897} & \multicolumn{1}{l}{0.19585} & \multicolumn{1}{l}{1.28502 } \\
	 \addlinespace
	\toprule
	\addlinespace
        Coefficients &  \multicolumn{1}{c}{Estimate} &  \multicolumn{1}{c}{Std. Error} &  \multicolumn{1}{c}{t}  &  \multicolumn{1}{c}{p} &Signif. \\
         \midrule
	intercept             & 1.3691  & 0.1361 &  10.056 & 7.5e-16 & *** \\
	impulse of push                 & 0.0057  & 0.0010 &  5.415 &6.26e-07 & *** \\
	distance none &         0.6052 &  0.2071 &  2.923 & 0.0045 &  ** \\  
	impulse : distance none & -0.0009 &  0.0013 & -0.668 & 0.5058 & \\   
	\midrule
	\addlinespace
	\multicolumn{5}{l}{Signif. codes: 0 ‘***’ 0.001 ‘**’ 0.01 ‘*’ 0.05 ‘.’ 0.1 ‘ ’ 1} \\
	\addlinespace
	\toprule
	\addlinespace
	\multicolumn{5}{l}{Residual standard error: 0.3636 on 80 degrees of freedom} \\
	\multicolumn{5}{l}{Multiple R-squared:  0.6621,	Adjusted R-squared:  0.6494 } \\
	\multicolumn{5}{l}{F-statistic: 52.25 on 3 and 80 DF,  p-value: $<$ 2.2e-16} \\
	\addlinespace
	\bottomrule
    \end{tabular}
\end{table}

\begin{table}[htbp]
\centering
    \caption{Results of the moderation analysis comparing no and elbow distance for A20.}
    \label{tab:statistics_a20}
    \begin{tabular}{lrrrll} 
    	\toprule
	\addlinespace
	 Residuals: &  \multicolumn{1}{l}{Min} &  \multicolumn{1}{l}{1Q} &  \multicolumn{1}{l}{Median} &  \multicolumn{1}{l}{3Q} &  \multicolumn{1}{l}{Max} \\
	 & \multicolumn{1}{l}{0.63112 } & \multicolumn{1}{l}{-0.15951} & \multicolumn{1}{l}{0.01043} & \multicolumn{1}{l}{0.16122} & \multicolumn{1}{l}{0.50499} \\
	 \addlinespace
	\toprule
	\addlinespace
        Coefficients &  \multicolumn{1}{c}{Estimate} &  \multicolumn{1}{c}{Std. Error} &  \multicolumn{1}{c}{t}  &  \multicolumn{1}{c}{p} &Signif. \\
         \midrule
	intercept             & 1.4637  & 0.0971 &  15.080 & $<$ 2e-16 & *** \\
	impulse of push                 & 0.0032  & 0.0007 &  4.326 & 4.34e-05 & *** \\
	distance none &         0.7921 &  0.1476 &  5.365 & 7.68e-07 &  *** \\  
	impulse : distance none & -0.0013 &  0.0010 & -1.338 & 0.185 & \\   
	\midrule
	\addlinespace
	\multicolumn{5}{l}{Signif. codes: 0 ‘***’ 0.001 ‘**’ 0.01 ‘*’ 0.05 ‘.’ 0.1 ‘ ’ 1} \\
	\addlinespace
	\toprule
	\addlinespace
	\multicolumn{5}{l}{Residual standard error: 0.2592 on 80 degrees of freedom} \\
	\multicolumn{5}{l}{Multiple R-squared:  0.7206,	Adjusted R-squared:  0.7101 } \\
	\multicolumn{5}{l}{F-statistic: 68.78 on 3 and 80 DF,  p-value: $<$ 2.2e-16} \\
	\addlinespace
	\bottomrule
    \end{tabular}
\end{table}

\begin{table}[ht]
\centering
    \caption{Results of the moderation analysis comparing no and elbow distance for formation B.}
    \label{tab:statistics_b}
    \begin{tabular}{lrrrll} 
    	\toprule
	\addlinespace
	 Residuals: &  \multicolumn{1}{l}{Min} &  \multicolumn{1}{l}{1Q} &  \multicolumn{1}{l}{Median} &  \multicolumn{1}{l}{3Q} &  \multicolumn{1}{l}{Max} \\
	 & \multicolumn{1}{l}{-0.55682} & \multicolumn{1}{l}{-0.21081} & \multicolumn{1}{l}{-0.00706} & \multicolumn{1}{l}{ 0.14132 } & \multicolumn{1}{l}{0.84558 } \\    
	 \addlinespace
	\toprule
	\addlinespace
        Coefficients &  \multicolumn{1}{c}{Estimate} &  \multicolumn{1}{c}{Std. Error} &  \multicolumn{1}{c}{t}  &  \multicolumn{1}{c}{p} &Signif. \\
         \midrule
	intercept            &1.6225&  0.1573&  10.314& 2.55e-13 &***\\
	impulse of push                 & 0.0042&  0.0011&   3.808& 0.0004 & *** \\
	distance none &         0.2405&  0.2285&   1.053& 0.2983 &  \\  
	impulse : distance none & 0.0010&  0.0015&   0.648& 0.5206   & \\   
	\midrule
	\addlinespace
	\multicolumn{5}{l}{Signif. codes: 0 ‘***’ 0.001 ‘**’ 0.01 ‘*’ 0.05 ‘.’ 0.1 ‘ ’ 1} \\
	\addlinespace
	\toprule
	\addlinespace
	\multicolumn{5}{l}{Residual standard error: 0.3218 on 44 degrees of freedom} \\
	\multicolumn{5}{l}{Multiple R-squared: 0.6176,	Adjusted R-squared:  0.5915  } \\
	\multicolumn{5}{l}{F-statistic: 23.69 on 3 and 44 DF,  p-value: 2.801e-09} \\
	\addlinespace
	\bottomrule
    \end{tabular}
\end{table}

\begin{table}[ht]
\centering
    \caption{Results of the moderation analysis comparing no and elbow distance for formation C.}
    \label{tab:statistics_c}
    \begin{tabular}{lrrrll} 
    	\toprule
	\addlinespace
	 Residuals: &  \multicolumn{1}{l}{Min} &  \multicolumn{1}{l}{1Q} &  \multicolumn{1}{l}{Median} &  \multicolumn{1}{l}{3Q} &  \multicolumn{1}{l}{Max} \\
	 & \multicolumn{1}{l}{-0.73434 } & \multicolumn{1}{l}{0.30283} & \multicolumn{1}{l}{-0.00189} & \multicolumn{1}{l}{0.25902} & \multicolumn{1}{l}{1.04705} \\
	 \addlinespace
	\toprule
	\addlinespace
        Coefficients &  \multicolumn{1}{c}{Estimate} &  \multicolumn{1}{c}{Std. Error} &  \multicolumn{1}{c}{t}  &  \multicolumn{1}{c}{p} &Signif. \\
         \midrule
	intercept             & 1.5211&   0.1973&   7.711& 1.05e-09 & *** \\
	impulse of push                 & 0.0049&   0.0015&   3.264&  0.0021  & ** \\
	distance none &         0.2523&   0.3088&   0.817&  0.4184&  \\  
	impulse : distance none & -0.0019&   0.0019&  -1.008&  0.3191    & \\   
	\midrule
	\addlinespace
	\multicolumn{5}{l}{Signif. codes: 0 ‘***’ 0.001 ‘**’ 0.01 ‘*’ 0.05 ‘.’ 0.1 ‘ ’ 1} \\
	\addlinespace
	\toprule
	\addlinespace
	\multicolumn{5}{l}{Residual standard error: 0.3975 on 44 degrees of freedom} \\
	\multicolumn{5}{l}{Multiple R-squared:  0.3343,	Adjusted R-squared:  0.289 } \\
	\multicolumn{5}{l}{F-statistic: 7.367 on 3 and 44 DF,  p-value: 0.000419} \\
	\addlinespace
	\bottomrule
    \end{tabular}
\end{table}

\newpage
In the one row experiments A5 and A20, a significant difference can be observed between no and elbow distance.
However, no significant distance was found for multiple rows B and C. 
\\

\newpage
In a next step, we compared formation A5 to formation A20 and the five-row experiments. 
The results are listed in Tables \ref{tab:comp_none} and \ref{tab:comp_elbow} for no and elbow distance respectively.

\begin{table}[htbp]
\centering
    \caption{Results of the moderation analysis comparing formations A5, A20 and five-row experiment at no distance.}
    \label{tab:comp_none}
    \begin{tabular}{lrrrll} 
    	\toprule
	\addlinespace
	 Residuals: &  \multicolumn{1}{l}{Min} &  \multicolumn{1}{l}{1Q} &  \multicolumn{1}{l}{Median} &  \multicolumn{1}{l}{3Q} &  \multicolumn{1}{l}{Max} \\
	 & \multicolumn{1}{l}{-0.78745 } & \multicolumn{1}{l}{-0.22060 } & \multicolumn{1}{l}{-0.03223} & \multicolumn{1}{l}{0.18316} & \multicolumn{1}{l}{1.28502} \\
	 \addlinespace
	\toprule
	\addlinespace
        Coefficients &  \multicolumn{1}{c}{Estimate} &  \multicolumn{1}{c}{Std. Error} &  \multicolumn{1}{c}{t}  &  \multicolumn{1}{c}{p} &Signif. \\
         \midrule
	intercept             & 1.9744&  0.1444&  13.671 & $<$ 2e-16 & *** \\
	impulse of push                 & 0.0048&  0.0008&   6.213 &1.74e-08& *** \\
	formation A20 &         0.2814&  0.2042&   1.378 &  0.1718   &  \\  
	formation five-row &        0.0140&  0.2624&   0.053  & 0.9576    &  \\ 
	impulse : formation A20 & -0.0028&  0.0011&  -2.595&   0.0111& *   \\ 
	impulse : formation five-row & 0.0020&  0.0021&   0.942 &  0.3489 & \\ 
	\midrule
	\addlinespace
	\multicolumn{5}{l}{Signif. codes: 0 ‘***’ 0.001 ‘**’ 0.01 ‘*’ 0.05 ‘.’ 0.1 ‘ ’ 1} \\
	\addlinespace
	\toprule
	\addlinespace
	\multicolumn{5}{l}{Residual standard error: 0.3365 on 87 degrees of freedom} \\
	\multicolumn{5}{l}{Multiple R-squared:  0.4297,	Adjusted R-squared:  0.397  } \\
	\multicolumn{5}{l}{F-statistic: 13.11 on 5 and 87 DF,  p-value: 1.626e-09} \\
	\addlinespace
	\bottomrule
    \end{tabular}
\end{table}

\begin{table}[htbp]
\centering
    \caption{Results of the moderation analysis comparing formations A5, A20 and five-row experiment at elbow distance.}
    \label{tab:comp_elbow}
    \begin{tabular}{lrrrll} 
    	\toprule
	\addlinespace
	 Residuals: &  \multicolumn{1}{l}{Min} &  \multicolumn{1}{l}{1Q} &  \multicolumn{1}{l}{Median} &  \multicolumn{1}{l}{3Q} &  \multicolumn{1}{l}{Max} \\
	 & \multicolumn{1}{l}{-0.66271} & \multicolumn{1}{l}{-0.13221} & \multicolumn{1}{l}{0.01054 } & \multicolumn{1}{l}{0.15693} & \multicolumn{1}{l}{0.75765} \\       
	 \addlinespace
	\toprule
	\addlinespace
        Coefficients &  \multicolumn{1}{c}{Estimate} &  \multicolumn{1}{c}{Std. Error} &  \multicolumn{1}{c}{t}  &  \multicolumn{1}{c}{p} &Signif. \\
         \midrule
	intercept             & 1.3691&  0.1108&  12.358 & $<$ 2e-16 & *** \\
	impulse of push                 & 0.0057&  0.0009&   6.654& 2.43e-09& *** \\
	formation A20 &        0.0946&  0.1567&   0.604  & 0.5476    &  \\  
	formation five-row &        -0.3296&  0.2178&  -1.514 &  0.1337       &  \\ 
	impulse : formation A20 & -0.0024&  0.0012&  -2.025 &  0.0459 & *   \\ 
	impulse : formation five-row & 0.0124&  0.0028&   4.412& 2.92e-05 & ***\\
	\midrule
	\addlinespace
	\multicolumn{5}{l}{Signif. codes: 0 ‘***’ 0.001 ‘**’ 0.01 ‘*’ 0.05 ‘.’ 0.1 ‘ ’ 1} \\
	\addlinespace
	\toprule
	\addlinespace
	\multicolumn{5}{l}{Residual standard error: 0.2959 on 87 degrees of freedom} \\
	\multicolumn{5}{l}{Multiple R-squared:  0.5715,	Adjusted R-squared:  0.5468 } \\
	\multicolumn{5}{l}{F-statistic: 23.2 on 5 and 87 DF,  p-value: 9.791e-15} \\
	\addlinespace
	\bottomrule
    \end{tabular}
\end{table}

\end{document}